\documentclass[12pt]{article}
\usepackage{amsmath}
\usepackage{amssymb}
\usepackage[mathscr]{eucal}
\usepackage[usenames]{color}
\usepackage[latin5]{inputenc}
\usepackage{url}
\definecolor{webgreen}{rgb}{0,0.5,0}

\newtheorem{prop}{Proposition}
\newtheorem{thm}{Theorem}

\numberwithin{equation}{section}
\numberwithin{thm}{section}
\numberwithin{lemma}{section}
\numberwithin{prop}{section}
\numberwithin{cor}{section}
\numberwithin{rmk}{section}
\numberwithin{defn}{section}

\newcommand{\gen}[1]{\partial_{#1}}

\newcommand{\curl}[1]{ \{#1\} }

\definecolor{darkolivegreen}{rgb}{0.333333, 0.419608, 0.1843140}

\DeclareMathOperator{\Sl}{sl}

\DeclareMathOperator{\PSL}{PSL}

\begin{document}
\pagenumbering{gobble}
\clearpage
\thispagestyle{empty}

\title{\Large Lie symmetries of a generalized Kuznetsov--Zabolotskaya--Khokhlov equation
}

\author{
F.~G\"ung\"{o}r\\ \small
Department of Mathematics, Faculty of Science and Letters,\\ \small Istanbul Technical University, 34469 Istanbul, Turkey \thanks{e-mail: gungorf@itu.edu.tr}\and
C. \"{O}zemir\\
\small Department of Mathematics, Faculty of Science and Letters,\\
\small Istanbul Technical University, 34469 Istanbul,
Turkey \thanks{e-mail: ozemir@itu.edu.tr} }

\date{}

\maketitle

\begin{abstract}
We consider a class of generalized Kuznetsov--Zabolotskaya--Khokhlov (gKZK) equations and determine its equivalence group, which is then used to give a complete symmetry classification of this class. The infinite-dimensional symmetry is used to reduce such equations to (1+1)-dimensional PDEs. Special attention is paid to group-theoretical properties of a class of generalized dispersionless KP (gdKP) or Zabolotskaya--Khokhlov equations as a subclass of gKZK equations. The conditions are determined under which a gdKP equation is invariant under a Lie algebra containing the Virasoro algebra as a subalgebra. This occurs if and only if this equation is completely integrable. A similar connection is shown to hold for generalized KP equations.
\end{abstract}

\section{Introduction}
The Kuznetsov--Zabolotskaya--Khokhlov equation (also known as 2+1 dimensional Burgers equation)
\begin{equation}\label{KZK}
(u_t+u u_x-u_{xx})_x- u_{yy}=0
\end{equation}
was originally derived in \cite{ZabolotskayaKhokhlov1969, Kuznetsov1971} as a model for the description of nonlinear acoustic beams. Later it has been used to deal with finite-amplitude compressional waves in solids. For several physical applications of \eqref{KZK} and its dispersionless version over the last two decades or so, see for example \cite{Rudenko2010, BakhvalovZhileikinRobert1987}.
A brief review of the works devoted to symmetries, Painlev\'e analysis, B\"acklund transformations and exact solutions can be found in Ref. \cite{Gungor2010a}.

A dispersionless version of \eqref{KZK} is the Zabolotskaya--Khokhlov equation, also called dispersionless Kadomtsev--Petviashvili (dKP) equation
\begin{equation}\label{ZK}
(u_t+u u_x)_x- u_{yy}=0
\end{equation}
which has first appeared in the study of unsteady motion in transonic flows, and then in the acoustic context.
Equation \eqref{ZK} is known to enjoy all the integrability properties of a typical integrable equation. It is integrable by inverse scattering technique, and has a Lax pair. It is also invariant under a Kac--Moody--Virasoro algebra isomorphic to that of KP equation \cite{Gungor2010a}. Some time ago, a variable coefficient generalization of \eqref{ZK} in the form
\begin{equation}\label{vcZK}
(u_t+p(t)u u_x)_x- u_{yy}=0
\end{equation}
was introduced in \cite{DunajskiPrzanowski} as the $\mathsf{U}(1)$-invariant null K\"ahler Einstein condition.   We shall show in Section 2 that it can be possible  to normalize  the coefficient $p(t)$ in \eqref{vcZK} to a constant, say 1 only when it  is a special power function. It also turns out that \eqref{vcZK} is completely integrable only for this very special value of the coefficient.

The purpose of this paper is to  carry out a  Lie symmetry analysis of a  generalized version  of \eqref{KZK} obtained by allowing its coefficients to be arbitrary functions of time in the form
\begin{equation}\label{vcBurgers}
(u_t+p(t)u u_x+q(t)u_{xx})_x+\sigma(t) u_{yy}=0,
\end{equation}
where $p$, $q$ and $\sigma$ are arbitrary nonzero functions. The case $q=0$  will   be investigated separately. This type of generalization can usually be considered as manifestation of a more realistic model  for variable geophysical conditions.
A study   towards analyzing symmetries and invariant solutions of a special case of \eqref{vcBurgers}, where only a single coefficient  is present,
\begin{equation}\label{refeq}
(u_t+u u_x-u_{xx})_x+\sigma(t) u_{yy}=0, \quad \sigma\ne 0
\end{equation}
appeared in Refs.~{\cite{Gungor2001, Gungor2001a}.
The drawback with these papers was the lack of an effective use of the notion of equivalence transformations. As a result, the list of equations obtained in the classification may contain some redundant classes which otherwise would not be so apparent to detect without recourse to them.   These transformations have the advantages of identifying equations with isomorphic symmetry algebras that can be transformed to each other. One of the motivations of this article is to emphasize the role they play in different applications, including picking out classes of variable coefficient equations  transformable to their constant coefficient counterparts \cite{VaneevaPopovychSophocleous2014}.

The organization of the article is as follows:  In Section \ref{section2}, we derive the equivalence group. In Section \ref{section3}, we use the standard algorithm to find the symmetry vector field  with coefficients satisfying an ODE   and then continue to determine the forms of symmetry algebras and the invariant equations for the gKZK equation of \eqref{vcBurgers}. We also look at group-theoretical properties of the generalized dKP (dispersionless Kadomtsev--Petviashvili) and the generalized KP equation. We specify the most general conditions for the equation to be invariant  under an arbitrary parametrization of time. We show that this can happen if and only if the equation can be transformed into the dKP equation itself. In Section \ref{section4}, a reduction to one-dimensional generalized (variable coefficient) Burgers equation is performed using the infinite-dimensional algebra as an application.

\section{Equivalence transformations}\label{section2}
Equivalence transformations have been used very effectively as an essential instrument in symmetry classification of differential equations, as well as discrete ones depending on unspecified functions of independent, dependent variables and their derivatives in general.  Lie group classification problems based on this approach  abound in the literature;  see for example \cite{Ovsiannikov1982} for prototype examples of several PDEs, \cite{GungorWinternitz2002, Gungor2010a, Basarab-HorwathGuengoerOezemir2013a, IvanovaSophocleousTracina2010} for applications to 2+1-dimensional variable coefficient evolution equations,   \cite{PocheketaPopovychVaneev2014, VaneevaPopovychSophocleous2014} for  1+1-dimensional general evolution equations and \cite{LeviWinternitzYamilov2011, LeviRiccaThomovaWinternitz2014} for generalized discrete differential-difference equations. An equivalence transformation   is one taking an equation from a given class of equations depending on arbitrary functions to another equation in the same class, but possibly with different arbitrary functions or coefficients here. The set of all  invertible smooth equivalence transformations  forms a Lie (pseudo-)group called the equivalence group  of the equation. Symmetry is an equivalence transformation preserving also the arbitrary functions. There are also point transformations that connect two fixed equations from a given class and are known in the literature under the names of allowed \cite{WinternitzGazeau1992} or form-preserving  \cite{KingstonSophocleous1998} or admissible transformations \cite{PopovychKunzingerEshraghi2010}. Classes of differential equations for which all these types of transformations are generated by equivalence ones are called normalized. (A detailed discussion of this notion can be found in \cite{PopovychKunzingerEshraghi2010}.)

Basically, there are two different approaches for finding  equivalence group. The first one is the infinitesimal method. Once a Lie algebra of equivalence vector fields acting not only on independent and dependent variables but also on parameter (coefficient) functions as new dependent variables has been found by the usual prolongation process, it can be integrated to  a local equivalence group. The second  one is the direct or global method in which the equivalence group is searched by applying the most general local point transformations
\begin{subequations}\label{1.2}
\begin{equation}\label{1.2a}
u=U(\tilde{x}, \tilde{y}, \tilde{t},\tilde u),
\quad x=X(\tilde x,\tilde y, \tilde t, \tilde u),
\quad y=Y(\tilde x,\tilde y, \tilde t, \tilde u),
\quad t=T(\tilde x,\tilde y, \tilde t, \tilde u),
\end{equation}
with nonvanishing total Jacobian determinant
\begin{equation}\label{1.2b}
J=\frac{D (X,Y,T,U)}{
D (\tilde x,\tilde y, \tilde t, \tilde u)}\ne 0,
\end{equation}
\end{subequations}
to the original equation and requiring that the transformed equation expressed in tildes should have exactly the same form as the initial one. The new coefficients in tildes will generally be different from the old ones induced by the transformations. When they happen to coincide ($\tilde{p}(\tilde{t})=p(\tilde{t})$, $\tilde{q}(\tilde{t})=q(\tilde{t})$, $\tilde{\sigma}(\tilde{t})=\sigma(\tilde{t})$) the equivalence transformations are nothing else but the Lie point symmetries.

The question of which method is more advantageous depends on the complexity of the structure of the equation. In our case, we find it more convenient to use the second method to construct the equivalence group of \eqref{vcBurgers}. It is easy to extract it from the results of Ref. \cite{Gungor2010a},  where a more general form of \eqref{vcBurgers} involving other derivatives with coefficients depending on a single variable $t$ or both $t$ and $y$ was studied. To this end, we simply specialize some coefficients to functions of $t$ alone, then the rest to zero and obtain the following result:
\begin{equation}\label{equiv}
\begin{array}{ll}
& u(x,y,t)=R(t)\tilde{u}(\tilde{x},\tilde{y},\tilde{t})-
\displaystyle\frac{\dot{\alpha}}{\alpha p}x+S(y,t),\\[.3cm]
& \tilde{x}=\alpha(t)x+\beta(y,t),\quad \tilde{y}=Y(y,t),\quad
\tilde{t}=T(t),\\[.3cm]
& \alpha\ne 0,\quad  R\ne 0, \quad Y_{y}\ne 0,\quad \dot{T}\ne
0,
\end{array}
\end{equation}
where the functions  $\alpha, \beta, Y, R, S$ must be chosen as consistent solutions of the system of PDEs
\begin{align}\label{comp}
\begin{aligned}
    \sigma Y_{yy}    &= 0   , \\
    2\sigma \beta _y Y_y  + \alpha Y_t &= 0    , \\
    \beta_t   \alpha  + pS\alpha^2  + \sigma \beta _y^2  &= 0 , \\
    - R\dot \alpha  + \dot R\alpha    + \sigma R\beta _{yy} &= 0  , \\
     - \frac{d}{dt}\left( {\frac{{\dot \alpha }}
{{\alpha p}}} \right)   + \frac{1}{p}\left( {\frac{{\dot \alpha }} {{\alpha
}}} \right)^2  + \sigma S_{yy} &= 0  ,
\end{aligned}
\end{align}
where the dots in \eqref{equiv}, \eqref{comp} and elsewhere will denote derivatives with respect to $t$.
In other words, transformations \eqref{equiv} take equations from the initial class \eqref{vcBurgers} to  those having  exactly the same differential structure but with the new coefficients (written in tildes). The nondegeneracy conditions in  \eqref{equiv} follow from the Jacobian requirement \eqref{1.2b}. Note that the transformations \eqref{equiv} are fiber-preserving or projectible because the new independent variables do not depend on $u$.
Here, the new coefficients in the transformed equation satisfy
\begin{equation}\label{coeff1}
\tilde p(\tilde t)   = p(t)\frac{{R\alpha }}
{{\dot T}},\qquad \tilde q(\tilde t) = q(t)\frac{{\alpha ^2 }}
{{\dot T}}, \qquad   \tilde \sigma (\tilde t)   = \sigma (t)\frac{{Y_y^2 }}
{{\alpha \dot T}}\;.
\end{equation}
We are now in a state to ask whether we can scale simultaneously $p$ and $q$ to 1 which is equivalent to requiring
$\tilde p(\tilde t) = 1$, $\tilde q(\tilde t) = 1$  under the transformation \eqref{equiv}. To achieve this normalization (scaling)  we  choose
the functions $R(t)$, $\alpha(t)$ and $T(t)$   in Eqs.
\eqref{equiv} so that for some $\delta\ne 0$ we have $Y_y=\delta(t)$ and
\begin{equation}\label{2.4}
\dot T(t) = q(t)\alpha ^2 (t),\qquad R(t) = \frac{q}
{p}\alpha.
\end{equation}
The remaining functions figuring in \eqref{equiv} must satisfy the compatibility equations \eqref{comp}. It is straightforward to see that there exists such a transformation with coefficients given by
\begin{align}\label{normal}
\begin{aligned}
\beta&=\frac{1}{2}\omega(t)y^2, & \omega(t)&=-\frac{\alpha}{\sigma}\frac{\dot{k}}{k},\\
R&=k(t)\alpha(t), & k(t)&=\frac{q}{p}, \quad  \dot{k}\ne 0,\\
Y&=\delta(t)y, & \delta(t)&=k^{2}(t), \\
S&=-\frac{1}{p\alpha^2}(\frac{1}{2}\alpha\dot{\omega}+\sigma\omega^2)y^2, &
\end{aligned}
\end{align}
where $\alpha$ satisfies a second order nonlinear ODE
$$-p\frac{d}{dt}\left( {\frac{{\dot \alpha }}
{{\alpha p}}} \right)+\left( {\frac{{\dot \alpha }} {{\alpha
}}} \right)^2+\frac{d}{dt}\Bigl(\frac{\dot{k}}{k}\Bigr)+\frac{\dot{k}}{k}\Bigl(\frac{\dot{\alpha}}{\alpha}-\frac{\dot{\sigma}}{\sigma}\Bigr)-2\Bigl(\frac{\dot{k}}{k}\Bigr)^2 =0.$$ Setting $A(t)=\dot{\alpha}/\alpha$, this ODE is reduced to a Riccati equation
\begin{equation}\label{Riccati}
\dot{A}=A^2+\frac{\dot{q}}{q}A+\frac{d}{dt}\Bigr(\frac{\dot{k}}{k}\Bigl)-2\Bigl(\frac{\dot{k}}{k}\Bigr)^2-\frac{\dot{\sigma}}{\sigma}\Bigl(\frac{\dot{k}}{k }\Bigr).
\end{equation}
Furthermore, $T$ is obtained from the integration
$$ T(t)=\int\alpha^2(t) q(t) \;dt, $$
and the action of the obtained equivalence transformation on $\sigma$ is given by
$$\tilde{\sigma} (\tilde t)   = \frac{\delta^2}{q\alpha^3}\sigma =\frac{q^3}{p^4 \alpha^3}\sigma.$$
When $\dot{k}= 0$, i.e. $p$ and $q$ are proportional, a reparametrization of time $t\to T(t)$ only will suffice for normalization and we have
$\tilde{\sigma}=q^{-1}\sigma$.

Under \eqref{equiv} together with \eqref{normal} we obtain the following canonical form of our initial class \eqref{vcBurgers}
\begin{equation}\label{main}
(u_t+u u_x+u_{xx})_x+\sigma(t) u_{yy}=0, \quad \sigma\ne 0.
\end{equation}
From now on we will restrict our analysis to the subclass \eqref{main}.
The equivalence transformations of \eqref{main} are obtained now by setting $p=q=1$ in \eqref{comp} and solving for $\alpha$, $\beta$, $T$, $Y$ and $S$ again. First of all, from the normalization condition we must have $R(t)=\alpha(t)$ and $\dot{T}=\alpha^2$. We then solve equations \eqref{comp} and find
\begin{subequations}\label{transfunct}
\begin{align}
Y(y,t)&=Y_1y+Y_0(t),\\
\beta(y,t)&=-\frac{\alpha\dot Y_0}{2Y_1\sigma}\,y+\beta_0(t),\\
S(y,t)&=\frac{1}{2Y_1}\Big(\frac{\ddot
Y_0}{\sigma}+\frac{\dot\alpha\dot Y_0}{\alpha \sigma}-\frac{\dot
Y_0\dot \sigma}{\sigma^2}\Big)\,
y-\frac{\dot\beta_0}{\alpha}-\frac{1}{4Y_1^2}\frac{\dot
Y_0^2}{\sigma},
\end{align}
\end{subequations}
where $Y_1\ne 0$ is a constant, $Y_0(t)$, $\beta_0(t)$ are arbitrary
functions.  Equation \eqref{Riccati} is simplified to the elementary equation $\dot{A}=A^2$ from which $\alpha$ and hence $T$ are obtained explicitly.
Alternatively, $T$
satisfies the special Schwarzian equation
$$ \frac{\dddot{T}}{\dot{T}}-\frac{3}{2}\Bigl ( \frac{\ddot{T}}{\dot{T}}\Bigr)^2=0,$$ whose solutions generate  M\"{o}bius transformations: $t\to (at+b)/(ct+d)$, where $\Delta=ad-bc\ne 0$. These transformations form a group isomorphic to $\PSL(2,\mathbb{R})$. That is why we  must have $\Delta>0$.

Summing up, we obtain
\begin{prop}
The  equivalence group $\mathsf{G_{E}}$ of the canonical  class \eqref{main} is infinite-dimensional and
given by
\begin{equation}\label{equiv2}
\begin{split}
\tilde{t}&=T(t)=\frac{at+b}{ct+d},\\
\tilde{x}&=\frac{\varepsilon\sqrt{\Delta}}{ct+d}\,x-\frac{\varepsilon\sqrt{\Delta}}{2Y_1}\frac{\dot
Y_0}{(ct+d)\sigma}\,y+\beta_0(t),\\
\tilde{y}&=Y_1y+Y_0(t),\\
u&=\frac{\varepsilon\sqrt{\Delta}}{ct+d}\,\tilde{u}+\frac{c}{ct+d}\,x+\frac{1}{2Y_1}
         \Big(\frac{\ddot Y_0}{\sigma}-\frac{c}{ct+d}\frac{\dot Y_0}{\sigma}-\frac{\dot Y_0\dot
         \sigma}{\sigma^2}\Big)\,y
         -\frac{1}{4Y_1^2}\frac{\dot
         Y_0^2}{\sigma}-\frac{\varepsilon}{\sqrt{\Delta}}(ct+d)\dot \beta_0,\\
\tilde{\sigma}&=\varepsilon\Delta^{-3/2}Y_1^2(ct+d)^3\sigma,
\end{split}
\end{equation}
where  $\varepsilon=\pm 1$, $a$, $b$, $c$, $d$ are arbitrary constants,  $\Delta=ad-bc>0$,  $Y_1$ is a nonzero constant and $Y_0(t)$, $\beta_0(t)$ are arbitrary
functions.

\end{prop}
{\bf Remarks:}
\begin{enumerate}
\item We note that the equivalence group  is maximal in the sense that all point transformations between equations from class \eqref{main} are exhausted by those from the group $\mathsf{G_{E}}$.
\item The quadruple of real numbers $(a,b,c,d)$ in \eqref{equiv2} is defined up to a nonzero multiplier so that one can set $\Delta=1$ with no loss of generality.
\item The $u$-component of transformations from~$\mathsf{G_E}$ is parameterized by the arbitrary element~$\sigma$.
The condition $\dot Y_0=0$ singles out those transformations from~$\mathsf{G_E}$ whose components associated with the dependent and the independent variables do not involve ~$\sigma$.
Therefore, these transformations constitute the usual equivalence group of the subclass (2.8).
\item   The old and new coefficients transform according to the relation
\begin{equation}\label{sigmatrans}
\sigma({t})=\varepsilon\Delta^{3/2}Y_1^{-2}(c{t}+d)^{-3}\tilde{\sigma}\Bigl(\frac{a{t}+b}{c{t}+d}\Bigr).
\end{equation}
An immediate conclusion of this relation is that the most general equation of the form \eqref{main} that can be transformed to its constant coefficient version should have the coefficient
$$\sigma=\sigma_0(t+\kappa)^{-3}$$  for some  constants $\sigma_0\ne 0$ and $\kappa$.
\end{enumerate}

The special case of \eqref{vcBurgers} for $q=0$ (generalized dKP equation) can be  treated by a si\-mi\-lar argument.
From the second condition of \eqref{coeff1} we have $\tilde{q}= 0$.
We try to solve the system \eqref{comp} for the functions figuring in \eqref{equiv}  such that  $\tilde{p}=1$, $\tilde{\sigma}=1$.
We find that this is possible only if the functions $p$ and $\sigma$ are related by
\begin{equation}\label{condstostandard}
p(t)=[c_1\int \sigma(t)\;dt+c_2]^{-3/2}\sigma(t),
\end{equation}
where $c_1$, $c_2$ are constants. The relevant functions in the equivalence transformations are given by
\begin{equation}\label{transtostandard}
\begin{split}
R&=[c_1\int \sigma(t)\;dt+c_2]^{-1/2}, \quad \alpha=1, \\
Y&=\delta(t)y+\nu(t), \quad \delta(t)=\delta_0 R^2,\\
\beta&=-\frac{\dot{R}}{2\sigma R}y^2+\beta_1(t)y+\beta_0(t),\\
S&=-\frac{1}{p}[(\dot{\beta}_1+\frac{\dot{\nu}\dot R}{\sigma \delta R})y+\dot{\beta}_0+\sigma\beta_1^2],\quad \beta_1=-\frac{\dot{\nu}}{2\sigma\delta},
\end{split}
\end{equation}
where   $\delta_0$ is a constant, and $\beta_0$, $\nu$ are arbitrary functions. $T(t)$ is obtained from the relation $\dot{T}=pR$.
Unless $c_1=0$,  $p$ and $\sigma$ are linearly independent.
For instance,  in the particular case $\sigma=1$, we have $p=p_0(t-t_0)^{-3/2}$, $p_0\ne 0$ and find that
\begin{equation}\label{svcZK}
(u_t+p_0 (t-t_0)^{-3/2}u u_x)_x+ u_{yy}=0
\end{equation}
is equivalent to its constant-coefficient form.

\section{Lie point symmetries}\label{section3}
\subsection{Symmetries of the gKZK equation}
First of all, we intend to perform a complete group classification of \eqref{main} for arbitrary given coefficient $\sigma(t)$.
A general element of the symmetry algebra of  \eqref{main} will have the form
\begin{gather} \label{VF}
V = \tau(x,y,t,u) \partial_t + \xi(x,y,t,u) \partial_x + \eta(x,y,t,u) \partial_y + \phi(x,y,t,u) \partial_u.
\end{gather}
From \eqref{equiv2} one can actually  assume a priori that $\tau=\tau(t)$, $\xi$ is linear in $x$, $\eta$ in $y$ and $\phi$ in $u$.
The standard infinitesimal symmetry requirement which is expressed as the annihilation of  (\ref{main}) on the solution set by the third prolongation ${\rm pr}^{(3)} V$ of (\ref{VF})
gives
the  determining equations for the coefficients $\tau$, $\xi$, $\eta$ and $\phi$. The solutions of the elementary ones among them are obtained as
$$\tau=\tau(t), \quad \xi=\frac{1}{2}\dot{\tau}x+\theta(y,t), \quad \eta=\eta(y,t), \quad \phi=-\frac{1}{2}\dot{\tau} u+\frac{1}{2}\ddot{\tau}x+\theta_{t}$$ with the following constraints (determining equations)
\begin{subequations}\label{deteqs}
\begin{align}
& \sigma \theta_{yy}=0,\qquad \sigma \eta_{yy}=0,\\
& \eta_t+2\sigma \theta_y=0,  \\
& \dddot{\tau}+2\sigma \theta_y=0, \\
& \tau \dot{\sigma}+[\frac{3}{2}\dot{\tau}-2\eta_y]\sigma=0 \label{last}.
\end{align}
\end{subequations}

Solving the determining equations \eqref{deteqs}, except  the last one we obtain
\begin{equation}\label{symVF}
V=\tau(t)\gen t+\bigl(\frac{1}{2}\dot{\tau}x+\theta(y,t) \bigr)\gen x+\eta(y,t)\gen y+[-\frac{1}{2}\dot{\tau}u+\frac{1}{2}\ddot{\tau}x+\theta_t]\gen u,
\end{equation}
where
\begin{equation}\label{coeff2}
\tau(t)=\tau_2 t^2 + \tau_1 t + \tau_0,\quad \theta(y,t)=-\frac{\dot{\eta_0}}{2\sigma}y+\xi_0(t),  \quad \eta(y,t)=\eta_1 y+\eta_0(t).
\end{equation}
In \eqref{coeff2}, $\tau_0$, $\tau_1$, $\tau_2$ and $ \eta_1 $ are arbitrary constants and $\xi_0$, $\eta_0$ are arbitrary functions.
The remaining determining equation \eqref{last} implies that the function $\sigma(t)$  must satisfy a first order ODE
\begin{gather} \label{classODE}
\big(\tau_2 t^2 + \tau_1 t + \tau_0\big) \dot{\sigma} + \left[3 \tau_2 t +\frac{3\tau_1}{2}-2\eta_1 \right] \sigma =0.
\end{gather}
The fact that \eqref{classODE} does not contain the functions $\xi_0$, $\eta_0$ suggests that the Lie point symmetry algebra is infinite-dimensional for any $\sigma$.
Indeed, splitting \eqref{classODE} with respect to $\sigma$ and $\dot{\sigma}$ gives $\tau=0$ ($\tau_2=\tau_1=\tau_0=0$) and $\eta_1=0$. So we find that the infinite-dimensional symmetry algebra  for  $\sigma(t)$ arbitrary is represented by the vector field (with $f=\xi_0$, $g=\eta_0$)
\begin{subequations}
\begin{equation}\label{V}
V=X(f)+Y(g),
\end{equation}
\begin{eqnarray}\label{A}
& X(f)&=f(t)\gen x+\dot{f}(t)\gen u, \\
& Y(g)&=g(t)\gen y-\frac{\dot{g}(t)}{2\sigma(t)}y\gen x-\frac{d}{dt}
\Bigl(\frac{\dot{g}(t)}{2\sigma(t)}\Bigr)y\gen u,
\end{eqnarray}
\end{subequations}
where $f(t)$ and $g(t)$ are arbitrary smooth functions.
The commutation relations are
\begin{equation}\label{comm1}
\begin{array}{ll}
&[X(f_1),X(f_2)]=0,\qquad [X(f),Y(g)]=0,\\[.5cm]
&[Y(g_1),Y(g_2)]=X\bigl(\displaystyle\frac{1}{2\sigma}(g_1' g_2-g_1 g_2')\bigr).
\end{array}
\end{equation}
The symmetry algebra is an infinite-dimensional nilpotent Lie algebra which we denote by
\begin{equation}\label{Linf}
L=\curl{X(f),Y(g)}.
\end{equation}

Now we identify all possible cases when the symmetry algebra is larger than \eqref{Linf}. We shall see that the algebra is extended at least by one additional basis element for other solutions ($\tau\ne 0$) of \eqref{classODE}.  One approach is to use an analysis given, for example in a recent paper \cite{VaneevaPopovychSophocleous2012}. The classifying ODE \eqref{classODE} has the structure
\begin{equation}\label{classODE2}
(pt^2+qt+r)\dot{\sigma}+(3pt+s)\sigma=0
\end{equation}
for some parameters $p,q,r,s$.
Under the action of the M\"{o}bius transformations in $t$ together with \eqref{sigmatrans}, eq. \eqref{classODE2} remains form-invariant with the new coefficients
\begin{align}
 \begin{aligned}
\tilde p&=pd^2-qcd+rc^2, &\quad
\tilde q&=-2pbd+q(ad+bc)-2rac,\\[1ex]
\tilde r&=pb^2-qab+ra^2, & \quad\tilde s&=\Delta s-3(pbd-qbc+rac).
 \end{aligned}
 \end{align}
These relations are defined up to a nonzero factor and the discriminant $D=\tau_1^2-4\tau_0\tau_2$ of the quadratic polynomial $\tau$ is a relative invariant, i.e. $\tilde{D}=\tilde{\tau}_1^2-4\tilde{\tau}_0\tilde{\tau}_2=\Delta^2 D$.
The existence of such an invariant suggests that, with an appropriate choice of the linearly independent pairs $(a,b)$ and $(c,d)$,   the triple $(p,q,r)$ can be transformed into one of the three inequivalent values depending on the sign of $D$. The remaining parameter $s$ can  be rescaled to $\pm 1$ only when $D=0$. The final list of inequivalent quadruples $(p,q,r,s)$ is given by
\[\left\{ {\begin{array}{*{20}{l}}
{(0,1,0,s_1 ),}&{D > 0,}\\
{(0,0,1,-1),}&{D = 0,}\\
{(1,0,1,s_2 ),}&{D < 0,}
\end{array}} \right.\]
where $s_1$ and $s_2\geq 0$ are  constants.
Once the simplified values of the parameters $(p,q,r,s)$ have been obtained, the corresponding ODE \eqref{classODE2} is solved in each case. We then substitute them into \eqref{classODE} and solve for the constants $\tau_0, \tau_1, \tau_2, \eta_1$ figuring in the symmetry algebra.

Alternatively, using the fact that the vector field
\begin{equation}\label{extendedsymVF}
\tilde{V}=(\tau_2 t^2 + \tau_1 t + \tau_0)\gen t+(\tau_2 t+\frac{\tau_1}{2})x \gen x+\eta_1y\gen y+[-(\tau_2 t+\frac{\tau_1}{2})u+\tau_2x]\gen u,
\end{equation}
which will generate the additional symmetries is invariant under the equivalence transformations
\begin{equation}
\begin{split}
\tilde{t}&=\frac{at+b}{ct+d},\quad \tilde{x}=\frac{x}{ct+d}, \quad \tilde{y}=Y_1y,\\
u&=\frac{\tilde{u}}{ct+d}+\frac{cx}{ct+d},\quad ad-bc=1,\\
\end{split}
\end{equation}
we can simplify the symmetry vector field \eqref{extendedsymVF} and reobtain the same representative equations as above. The M\"{o}bius transformations act on the $t$-coefficient $\tau$ of the vector field $\tilde{V}$
by conjugation. This means we can transform $\tau\gen t$ which represents a general element of an $\Sl(2,\mathbb{R})$ algebra into one of its one-dimensional subalgebras: $t\gen t$, $\gen t$, $(t^2+1)\gen t$ depending on the sign of the adjoint action invariant $D=\tau_1^2-4\tau_0\tau_2$ being positive, zero and negative, respectively.

In the following we sum up the additional symmetries:

\begin{enumerate}

\item $D>0$:
We obtain
\begin{gather*} \label{c4}
\sigma(t) = \sigma_0 t^s.
\end{gather*}
$\sigma_0$ can be rescaled to $\pm 1$ by a scaling transformation in $y$.
The corresponding symmetry for $s\ne -3$ is
$$D(s)=2t\gen t+x\gen x+\frac{(2s+3)}{2} y\gen y-u\gen u.$$
In the special case $s=-3$, there are two additional symmetries
$$D(-3)=2t\gen t+x\gen x-\frac{3}{2} y\gen y-u\gen u, \quad C_0\equiv C(0)=t^2\gen t+xt\gen x+(x-tu)\gen u.$$
Note that under the  equivalence transformation
\begin{equation}\label{equiv4}
\tilde{t}=-\frac{1}{t}, \quad \tilde{x}=\frac{x}{t}, \quad \tilde{y}=y, \quad \tilde{u}=tu-x
\end{equation}
we have $s \rightarrow s+3$. This means that the case $s=-3$ is equivalent to $\sigma(t) = \sigma_0$.  Under the same transformation  $D(-3)$ and $C_0$ get transformed to $D(0)$ (up to sign) and $T_0=\gen t$, respectively.

\item $D=0$: We have $\sigma(t)=\sigma_0 e^t$, $\sigma_0=\pm 1$. The corresponding symmetry is
$$T=\gen t+\frac{1}{2}y\gen y.$$

\item $D<0$: We have
    \begin{gather*} \label{c1}
    \sigma(t)=\sigma_0\big(1+t^2\big)^{-3/2} e^{s \arctan t}, \qquad \sigma_0 = \pm 1.
    \end{gather*}
The additional symmetry is
$$ C(s)=(1+t^2)\gen t+xt\gen x+\frac{s}{2}y\gen y+(x-tu)\gen u.$$
\end{enumerate}

\noindent{\bf Remarks:}
\begin{enumerate}
\item Note that if $\sigma=\sigma_0=\pm 1$, then we have $\tau_2=0$, $\eta_1=\frac{3}{4}\tau_1$ and $\tau_0$ is arbitrary. This gives two additional symmetry generators
$$T_0=\gen t,\quad D_0\equiv D(0) =2t\gen t+x\gen x+\frac{3}{2}y\gen y-u\gen u.$$

\item All the extended algebras have a semi-direct sum structure
$$L=X_0\vartriangleright \curl{X(f),Y(g)},$$ where the ideal is on the right  and $X_0$ is one of the basis elements $C$, $T$,  and $D$. For $\sigma=\pm 1$, $X_0=\curl{T_0, D_0}$.

\end{enumerate}

\begin{table}[t]\centering
\caption{Symmetry classification of \eqref{main}; $\sigma_0 = \pm 1$.}\label{table1}
\vspace{1mm}

\begin{tabular}{|c|c|c|c|}
\hline
$N$ & $\sigma(t)$  & Basis elements of symmetry algebra $L$ \\ \hline \hline
1 & $\forall$ &  $X(f)$, $Y(g)$\\ \hline
2 & $\sigma_0$ & $X(f)$, $Y(g)$, $T_0$, $D_0$ \\ \hline
3 & $\sigma_0 t^s$  & $X(f)$, $Y(g)$, $D(s)$, $s\ne -3$ \\ \hline
4 & $\sigma_0 e^t$  & $X(f)$, $Y(g)$, $T$ \\ \hline
5 & $\sigma_0 (t^2+1)^{-3/2} e^{s \arctan t}$ &  $X(f)$, $Y(g)$, $C(s)$ \\ \hline
\end{tabular}
\end{table}
In Table~\ref{table1} we present representatives of all classes of functions~$\sigma(t)$ for the symmetry algebra.   The classification is done under the equivalence transformations~(\ref{equiv2}).

\subsection{KMV symmetries of the gdKP equation}\label{KMV-gdKP}
We now turn to study symmetries of the generalized dKP (gdKP) equation
\begin{equation}\label{gdKP}
(u_t+p(t)u u_x)_x+\sigma(t) u_{yy}=0, \quad p\cdot\sigma\ne 0.
\end{equation}
Allowing variable coefficients in a constant coefficient integrable PDE usually destroys its integrability. It is well-known that just as the usual KP equation,  its dispersionless variant dKP is also integrable and has an infinite-dimensional symmetry algebra with a specific loop structure \cite{Gungor2010a}. It would be  interesting to know the subcases   when the equation \eqref{gdKP} is integrable or at least transformable to the usual dKP equation by point transformations.  One can of course always scale  the coefficient $p\to 1$ by a reparametrization of time and then classify its symmetries. We do not choose this option. But rather, we are interested in looking at the possibility if the symmetry algebra of the gdKP equation can be invariant under an entire Kac--Moody--Virasoro (KMV) algebra.

The Lie infinitesimal process is applied to show that the symmetry algebra of \eqref{gdKP} can be represented by the vector field
\begin{equation}\label{dKPVF}
V=\tau(t)\gen t+[\mu(t)x+\theta(y,t)]\gen x+(\eta_1(t)y+\eta_0(t))\gen y+[Q(t)u+\frac{1}{p}(\dot{\mu}x+\theta_t)]\gen u,
\end{equation}
\begin{subequations}
\begin{eqnarray}
\mu(t)&=&\frac{1}{3}\dot{\tau}+\frac{\tau}{9}\Bigl(4\frac{\dot{p}}{p}-\frac{\dot{\sigma}}{\sigma}\Bigr)+C_0,\\
\eta_1(t)&=&\frac{2}{3}\dot{\tau}+\frac{2}{9}\tau\Bigl(\frac{\dot{p}}{p}+2\frac{\dot{\sigma}}{\sigma}\Bigr)+\frac{C_0}{2},\\
Q(t)&=&-\frac{2}{3}\dot{\tau}-\frac{\tau}{9}\Bigl(5\frac{\dot{p}}{p}+\frac{\dot{\sigma}}{\sigma}\Bigr)+C_0, \\
\theta(y,t)&=& \frac{1}{2\sigma}(\dot{\mu}+\dot{Q})y^2-\frac{\dot{\eta}_0(t)}{2\sigma}y+\theta_0(t),
\end{eqnarray}
where $C_0$ is a constant and $\eta_0$, $\theta_0$ are arbitrary functions.
The remaining determining equation for $\tau$ provides the classifying ODE
\begin{equation}\label{mueq}
3\ddot{\mu}-\Bigl(\frac{\dot{p}}{p}+2\frac{\dot{\sigma}}{\sigma}\Bigr)\dot{\mu}=-\frac{\dot{\sigma}}{\sigma}
\frac{d}{dt}\Bigl(\frac{1}{p}\frac{d}{dt}(p\tau)\Bigr)+\frac{d^2}{dt^2}\Bigl(\frac{1}{p}\frac{d}{dt}(p\tau)\Bigr).
\end{equation}
\end{subequations}
First of all, the parameter $C_0$ generates a dilational symmetry
$$D=2x\gen x+y\gen y+2u \gen u$$ for any $p$ and $\sigma$.
We are interested in specifying conditions  on  $p$, $\sigma$ that leave $\tau$ free. Eq. \eqref{mueq} can be viewed as a linear relation in $\tau$ and its derivatives with coefficients depending on $p$, $\sigma$ and derivatives. We note that the coefficients of $\dddot{\tau}$ and $\ddot{\tau}$ in \eqref{mueq} are identically zero and eq. \eqref{mueq} has the form $E(t)\dot{\tau}+F(t)\tau=0$, where, with $K=\dot{p}/p$, $L=\dot{\sigma}/\sigma$,
\begin{equation}\label{EF}
\begin{split}
E(t)&=6(\dot{K}-\dot{L})-4K^2+2KL+2L^2,\\
F(t)&=3(\ddot{K}-\ddot{L})-(4K-L)\dot{K}+(K+2L)\dot{L}.
\end{split}
\end{equation}
If we force $E(t)$ and $F(t)$ to vanish simultaneously we obtain the necessary conditions for $\tau$ figuring as $t$-coefficient in the vector field to remain free.  A compatibility  of the resulting equations    provides
\begin{equation}\label{comp2}
2\Bigl(\frac{\dot{p}}{p}\Bigr)^2-3\frac{d}{dt}\Bigl(\frac{\dot{p}}{p}\Bigr)-\frac{\dot{p}}{p}\frac{\dot{\sigma}}{\sigma}
+3\frac{d}{dt}\Bigl(\frac{\dot{\sigma}}{\sigma}\Bigr)-\Bigl(\frac{\dot{\sigma}}{\sigma}\Bigr)^2=0.
\end{equation}
This condition can be shown to be  equivalent  to the condition  \eqref{condstostandard} for the gdKP equation to be transformable to the standard dKP equation. Indeed, making the change of variable $p=\sigma e^{\Omega(t)}$ in \eqref{comp2} results in the ODE
$$\ddot\Omega-\frac{\dot
\sigma}{\sigma}\dot\Omega-\frac{2}{3}\dot\Omega^2=0,$$ which can be integrated to give
$$e^{\Omega}=\Big[c_1\int\sigma(t)dt+c_2\Big]^{-3/2}.$$
This is exactly the condition \eqref{condstostandard} relating $p$ and $\sigma$ when the equation can be transformed to \eqref{ZK} by a point transformation.
This implies that under condition \eqref{condstostandard} the vector field \eqref{dKPVF} depends on three arbitrary functions of time, and the general element of the symmetry algebra of the gdKP equation can be written as
\begin{subequations}\label{symVFgdKP}
\begin{equation}
V=T(\tau)+X(\theta_0)+Y(\eta_0)+\gamma D,
\end{equation}
where $\tau, \theta_0, \eta_0$ are arbitrary smooth functions of  $t$, $\gamma$ is a parameter. Here
\begin{align}
{T} (\tau) \;&=\; \tau \partial_t +(\mu x+\chi y^2) \partial_x + \eta_1 y\partial_y +
[Q u +\frac{1}{p}(\dot{\mu} x+\dot{\chi}y^2)] \partial_u,\\
{X}(\theta_0) \; &=\; \theta_0 \partial_x + \frac{1}{p}\dot{\theta}_0 \partial_u, \\
{Y}(\eta_0) \; &=\;  -\frac{1}{2\sigma} \dot{\eta}_0 y \partial_x + \eta_0\partial_y
-\frac{1}{p}  \frac{d}{dt}(\frac{\dot{\eta}_0}{2\sigma}) y\partial_u,\\
\chi&=\frac{1}{2\sigma}(\dot{\mu}+\dot{Q}). \nonumber
\end{align}
\end{subequations}
The basis elements  $\curl{T(\tau), X(\xi), Y(\eta), D}$ of the Lie symmetry algebra satisfy the commutation relations
\begin{subequations}\label{com}
\begin{alignat}{2}
[{T}(\tau_1), {T}(\tau_2)]\; &= \; {T}{ (\tau_1 \dot{\tau}_2
-\dot{\tau}_1 \tau_2 )}, \medspace & & \\
[D, {X}(\xi)]\;  &= \; -2 {X}(\xi),\medspace &
[{X}(\xi), {Y}(\eta)]\; &= \;
0,  \\
[D, {Y}(\eta)]\;  &= \; - {Y}(\eta),\medspace  &
[{X}(\xi),{T}(\tau)]\;
&= \; {X}{(\pi_x )} ,  \\
[D, {T}(\tau)]\; &= \;0, \medspace & [{Y}(\eta), {T}(\tau)]\; &= \; {Y}{ (\pi_y)},  \\
[{X}(\xi_1), {X}(\xi_2)] \; &= \; 0,  \medspace &  [{Y}(\eta_1), {Y}(\eta_2)] \; &= \; {X}\Bigl(-{\frac{1}{2\sigma} (\eta_1 \dot{\eta}_2- \dot{\eta}_1 \eta_2 )\Bigr)},
\end{alignat}
where
$$\pi_x =\frac{1}{3}\dot{\tau} \xi - \tau \dot{\xi}+\frac{1}{9}\xi\tau\Bigr(4\frac{\dot{p}}{p}-\frac{\dot{\sigma}}{\sigma}\Bigr),\quad \pi_y =\frac{2}{3} \dot{\tau}\eta - \tau \dot{\eta}+\frac{2}{9}\eta \tau\Bigr(\frac{\dot{p}}{p}+2\frac{\dot{\sigma}}{\sigma}\Bigr).$$
\end{subequations}
The commutation relations suggest that the infinite-dimensional symmetry algebra $L$ of the gdKP equations has the structure of a Kac--Moody--Virasoro (KMV) algebra  only if the condition \eqref{condstostandard} is satisfied among $p$ and $\sigma$. This is seen by introducing a basis for the functions $\tau(t)$, $\xi(t)$, $\eta(t)$ consisting of the monomials $t^n, n\in \mathbb{Z}$ (in other words, restricting them to be Laurent polynomials in $t$) and obtaining a basis $\curl{T(t^n), X(t^n), Y(t^n)}$ for the corresponding Lie algebra  having exactly the same commutation relations with those of a (centerless) KMV type algebra. (See \cite{DavidKamranLeviWinternitz1986} for more details.) This fact also indicates that an integrable subclass, which we call the integrable gdKP equation, has been singled out of the original gdKP class.  The symmetry algebra of the integrable gdKP equation can be written as a semidirect sum
\begin{equation}\label{KMV}
L=\curl{T(\tau)}\rhd \curl{X(\xi), Y(\eta), D}
\end{equation}
with the ideal  on the right.
We sum up the result as two theorems.
\begin{thm}\label{t1}
The generalized dKP (gdKP) equation  \eqref{gdKP} admits the (centerless) Virasoro algebra as
a symmetry algebra if and only if the coefficients satisfy \eqref{condstostandard}.
\end{thm}

\begin{thm}\label{t2}
The gdKP equation \eqref{gdKP} is invariant under a Lie
symmetry  algebra  which contains a Virasoro
algebra as a subalgebra, if and only if it can be transformed into
the dKP equation itself by a point transformation. The transformation is given by \eqref{transtostandard}.
\end{thm}

If $E(t)\neq 0$ in $E(t)\dot{\tau}+F(t)\tau=0$, then this first order ODE  fixes $\tau=\hat{\tau}=\tau_0\exp\curl{\int(-F(t)/E(t)dt)}\equiv \tau_0 H(t)$  and the loop structure of the algebra is lost though it is still infinite-dimensional.  The Virasoro part $\curl{T(\tau)}$ of the Lie symmetry algebra \eqref{KMV} is no longer present for all other values of $p$ and $\sigma$ where $E(t)\cdot F(t)\neq 0$. In this case, the symmetry element $T(\hat{\tau})$ is all that is remnant of  the arbitrary reparametrization of time corresponding to $T(\tau)$. The algebra gets largely reduced. A basis for the algebra is $\curl{X(\xi), Y(\eta), T(H), D}$. The subalgebra $\curl{X(\xi), Y(\eta)}$ is a centerless Kac--Moody algebra. If $E(t)\neq 0$, but $F(t)\equiv0$, then $\tau=\tau_0$, a constant. This happens when $p(t)=Ae^{kt}$ and $\sigma(t)=Be^{\ell t}$ for $k/\ell \ne (1\pm \sqrt{3})/2$. The  group transformations generated by $T(\tau_0)$ are only  a time translation plus dilations in the other variables.

\subsubsection{Special integrable subcases of \eqref{gdKP}} We now consider some special cases satisfying condition \eqref{condstostandard}.
\begin{enumerate}
\item $p=1$ and $\sigma=1$ (the standard dKP equation): Eq. \eqref{comp2} is trivially satisfied and we have  $\mu=\dot{\tau}/3$ (ignoring  constant $C_0$) and  $Q=-2\dot{\tau}/3$.  So the symmetry algebra of dKP equation is recovered as
$$V=T(\tau)+X(\xi)+Y(\eta),$$
\begin{subequations}\label{ZKSYM}
\begin{align}
{T} (\tau) \;&=\; \tau \partial_t + \frac{1}{6} \left(2\dot{\tau} x  -\ddot{\tau} y^2
\right) \partial_x + \frac{2  \dot{\tau}}{3} y \partial_y + \frac{1}{6}
\left(- 4\dot{\tau} u  + 2 \ddot{\tau} x  -  \dddot{\tau} y^2\right) \partial_u,\\
{X}(\xi) \; &=\; \xi \partial_x + \dot{\xi} \partial_u, \\
{Y}(\eta) \; &=\;  -\frac{1}{2}  \dot{\eta} y\partial_x + \eta\partial_y
-\frac{1}{2}  \ddot{\eta} y\partial_u,
\end{align}
\end{subequations}
where $\tau, \xi, \eta$ are arbitrary smooth functions of  $t$.
\item $p=1$, $\sigma=t^{-3}$ (up to translation and scaling of integration constants): The symmetry vector field
\eqref{symVFgdKP} has the coefficients
$$\mu(t)=\frac{1}{3}(\dot{\tau}+\frac{\tau}{t}),\quad \eta_1(t)=\frac{2}{3}(\dot{\tau}-2\frac{\tau}{t}),\quad Q(t)=-\frac{1}{3}(2\dot{\tau}-\frac{\tau}{t}).$$
\item $p=t^{-3/2}$, $\sigma=1$: We have
$$\mu(t)=\frac{1}{3}(\dot{\tau}-2\frac{\tau}{t}),\quad \eta_1(t)=\frac{1}{3}(2\dot{\tau}-\frac{\tau}{t}),\quad Q(t)=-\frac{1}{3}(2\dot{\tau}-\frac{5}{2}\frac{\tau}{t}).$$
\end{enumerate}
The  point transformations taking the corresponding gdKP equations to the standard form are given by \eqref{transtostandard}.

\subsection{KMV symmetries of the gKP equation}\label{KMV-gKP}
An analogous situation occurs for the generalized KP (gKP) equation
\begin{equation}\label{gKP}
(u_t+p(t)uu_x+q(t)u_{xxx})_x+\sigma(t)u_{yy}=0,
\end{equation}
where $p$, $q$, $\sigma$ are arbitrary nonzero functions. The equivalence and symmetry determining equations for a more general form of this equation were obtained in \cite{GungorWinternitz2004}. Using the results of \cite{GungorWinternitz2004}, one can show that the following   are necessary and sufficient conditions for the existence of the centerless KMV loop algebra as a symmetry algebra, passing the Painlev\'e test and transformability to the standard form:
\begin{equation}\label{condstostandard2}
q(t)=p(t)[c_1\int p(t)dt+c_2], \quad \sigma(t)=Y_1^{-2}p(t)[c_1\int p(t)dt+c_2]^{-3},
\end{equation}
where $c_1$, $c_2$ ($c_1. c_2\ne 0$), $Y_1\ne 0$  are constants.
The point transformation taking \eqref{gKP} to itself for $p=q=\sigma=1$ (normalizing  transformation) is
\begin{equation}\label{equiv3}
\begin{array}{ll}
& u=\alpha(t)\tilde{u}-
\displaystyle\frac{\dot{\alpha}}{\alpha p}x-\frac{1}{p\alpha^2}[\alpha \beta_t+\sigma \beta_y^2],\\[.3cm]
& \tilde{x}=\alpha(t)x+\beta(y,t),\quad \tilde{y}=Y_1y+Y_0(t),\\[.3cm]
&\tilde{t}=\int p\alpha^2dt, \quad \beta(y,t)=-\dfrac{\alpha}{2\sigma}\dot{Y}_0y+\beta_0(t),\\[.3cm]
&\alpha=[c_1\int p(t)dt+c_2]^{-1},
\end{array}
\end{equation}
where $Y_0$, $\beta_0$ are arbitrary functions. Two equations lying in the subclass \eqref{condstostandard2} are $(p,q,\sigma)=(1,t,t^{-3})$ and $(p,q,\sigma)=(e^t,e^{2t},e^{-2t})$. A normalizing point transformation $(p,q,\sigma)\to (1,1,1)$ in the first case is given by \eqref{equiv4}. In the second one, one such transformation is
$$\tilde{t}=e^{-t}, \quad \tilde{x}=e^{-t}x, \quad \tilde{y}=y, \quad \tilde{u}=e^{t}u-x.$$

We now look at the symmetry properties of the gKP equation (the normalization of the coefficient $p$ is again ignored for our purposes). Our intention,  as in the gdKP equation, is to  find  the conditions on $p$, $q$ and $\sigma$ keeping intact the KMV structure of the symmetry algebra of the standard KP equation. The vector field representing the symmetry algebra of \eqref{gKP} \cite{GungorWinternitz2004} has the form
\begin{equation}\label{KPVF}
V=\tau(t)\gen t+[a(t)x+b(y,t)]\gen x+[\eta_1(t)y+\eta_0(t)]\gen y+[R(t)u+\frac{1}{p}(\dot{a}x+b_t)]\gen u,
\end{equation}
\begin{subequations}
\begin{eqnarray}
a(t)&=&\frac{1}{3}(\dot{\tau}+L \tau),\quad \eta_1(t)=\frac{1}{2}(a+\dot{\tau}+M \tau),\\
R(t)&=&-\frac{2}{3}\dot{\tau}+(\frac{1}{3}L-K)\tau, \\
b(y,t)&=& -\frac{1}{8\sigma}[\dot{a}+\ddot{\tau}+\frac{d}{dt}(M \tau)]y^2-\frac{\dot{\eta}_0(t)}{2\sigma}y+\xi_0(t),
\end{eqnarray}
where $K=\dot{p}/p$, $L=\dot{q}/q$, $M=\dot{\sigma}/\sigma$ and $\xi_0$, $\eta_0$ are arbitrary functions. The determining equation for $\tau$ is
\begin{equation}
\begin{split}
(-4K+3L+M)\ddot{\tau}+(6\dot{L}-6\dot{M}-4KL+LM+3M^2)\dot{\tau}
\\+(3\ddot{L}-3\ddot{M}-4K\dot{L}+M \dot{L}+3M\dot{M})\tau=0.
\end{split}
\end{equation}
\end{subequations}
If $\tau$ is supposed to be arbitrary we must set the coefficients equal to zero. The equations obtained are consistent if we have, in terms of the coefficients,
$$4\frac{\dot{p}}{p}-3\frac{\dot{q}}{q}-\frac{\dot{\sigma}}{\sigma}=0, \qquad 2\frac{d}{dt}\Bigl(\frac{\dot{\sigma}}{\sigma}-\frac{\dot{q}}{q}\Bigr)+\Bigl(\frac{\dot{q}}{q}\Bigr)^2-
\Bigl(\frac{\dot{\sigma}}{\sigma}\Bigr)^2=0.$$
The above equations can be  integrated to give the conditions \eqref{condstostandard2}. This implies that the KMV structure can persist only for equations that are equivalent to the KP itself under the point transformation \eqref{equiv3}.

\section{Reduction of \eqref{main} to the one-dimensional Burgers equation}\label{section4}
Reductions by subalgebras $X(f)$ and $Y(g)$ for \eqref{main} in the generic case were presented in \cite{Gungor2001a}. Moreover, in the nongeneric case where $\sigma$ is  a power or an exponential function, reductions by two dimensional subalgebras of the symmetry algebra  to ODEs and analysis of their possible solutions that led to exact solutions of the original PDE were investigated in the same work. Here we would like to give reductions in the generic case by the  algebra $L$ represented by $X(f)+Y(g)$, $g\ne 0$ which is actually conjugate under the group  $\exp\curl{Y(G)}$ for some $G(t)$ to $Y(g)$.
An invariant solution  must then have the form
\begin{equation}
\begin{split}
&u=W(\xi,t)-\frac{1}{g}[\frac{d}{dt}\Bigl(\frac{\dot{g}}{4\sigma}\Bigr)y^2-\dot{f}y],\\
&\xi=x+\frac{\dot{g}}{4\sigma g}y^2-\frac{f}{g}y.
\end{split}
\end{equation}
Once integrated, the reduced equation satisfies
$$W_t+WW_{\xi}+W_{\xi\xi}+\sigma \Bigl(\frac{f}{g}\Bigr)^2 W_{\xi}+\frac{\dot{g}}{2g}W-\frac{\sigma}{2g}\frac{d}{dt}\Bigl(\frac{\dot{g}}{\sigma}\Bigr)\xi+\rho(t)=0,$$
where $\rho$ is an arbitrary function which can be transformed away by a  time dependent translation of the dependent variable $W \to W+h(t)$  with $h$ appropriately chosen (as a solution of a first order linear ODE).
Note that the above translation has an effect of increasing the coefficient of the first derivative term $W_{\xi}$ by $h$. $W_\xi$ can also be set to zero. This is achieved by the transformation
$$W=F(z,t), \quad z=\xi+\gamma(t),$$ where $\gamma$ satisfies
$$\dot{\gamma}=h-\Bigl(\frac{f}{g}\Bigr)^2\sigma.$$
Further simplification comes from the choice
$\dot{g}=\sigma$. The final form of the reduced equation can be written as
\begin{equation}\label{1-dim}
F_t+FF_z+F_{zz}+\frac{\dot{\Sigma}}{2\Sigma}F=0, \quad \dot{\Sigma}=\sigma,
\end{equation}
which is a one-dimensional generalized Burgers equation.
The transformation
$$\tilde{t}=T(t)=\int\Sigma^{-1/2}dt,  \quad \tilde{z}=z,  \quad F=\Sigma^{-1/2} \tilde{F}(\tilde{z},\tilde{t})$$
takes  equation \eqref{1-dim} to the more usual form
\begin{equation}\label{1-dim2}
\tilde{F}_{\tilde{t}}+\tilde{F}\tilde{F}_{\tilde{z}}+\tilde{g}(\tilde{t})\tilde{F}_{\tilde{z}\tilde{z}}=0,
\end{equation}
where
$\tilde{g}(\tilde{t})=\Sigma^{1/2}(\tilde{t})$.
A  group classification of \eqref{1-dim2}, completing the partial results in the literature,  appeared very recently in Ref. \cite{VaneevaSophocleousLeach2013}. According to the results of this paper, the maximal symmetry algebra of  \eqref{1-dim2},  which occurs when $g=\pm 1$ is five dimensional and isomorphic to the complete Galilei algebra (Galilei-similitude algebra extended by projective elements) having the semi-direct sum structure $\Sl(2,\mathbb{R})\vartriangleright 2 \mathsf{A}$. This is the usual non-potential Burgers equation well-known for its linearizability  to the heat equation by the Hopf--Cole transformation. For three special forms of $g$ it is three-dimensional, or otherwise  two-dimensional generating space translations and Galileian boosts  for arbitrary $g$.

\section{Conclusions}\label{section5}
We have demonstrated that equivalence transformations are useful in symmetry classification.  A knowledge of equivalence group enabled us to simplify the arbitrary coefficients figuring in Eq. \eqref{vcBurgers} (we have been able to normalize two out of three), then proceed to classify all equations with nontrivial symmetries. The equation under study is shown to have an infinite-dimensional symmetry algebra for arbitrary coefficient $\sigma$. The symmetry algebra is extended at least by one additional element when $\sigma$ has special forms. We again made use of equivalence transformations to give only a representative list of equations  which cannot be transformed into each other. The results of  this paper  reveal that some of the equations presented in the classification list of  Refs.~\cite{Gungor2001a, Gungor2001} are redundant. Also,  symmetry properties of the gdKP equation are studied.     This is precisely where the equivalence transformations come in again. Requiring that the equation be invariant under a reparametrization of time we have identified the most general equation of the form \eqref{gdKP} allowing a Kac--Moody--Virasoro algebra  and have shown that this equation can be transformed into  the integrable dKP equation. We shall call this class the integrable gdKP equation. Thus, all variable coefficient integrable nonlinear PDEs  are singled out from the generic  nonintegrable gdKP equation. A similar link between the generalized KP (gKP) and KP has also been emphasized.
Based on this fact  we can transform the Lax pair and any solution for the dKP (or KP) equation to a Lax pair and to a solution for the gdKP (or gKP) equation.  For a further discussion of this topic and other applications we refer the interested readers to a very recent work \cite{VaneevaPopovychSophocleous2014} for PDEs in 1+1-dimension and \cite{GungorWinternitz2002, LeviWinternitz1988} in 2+1-dimension. The relationship between infinite-dimensional Lie point symmetries of PDEs and integrability is not well understood so far. We hope that the  results of Subsections \eqref{KMV-gdKP} and \eqref{KMV-gKP} can shed some light on this issue.

Finally, to the best of our knowledge, the question of whether \eqref{condstostandard} is the only case when the gdKP equation passes the Painlev\'e test remains unanswered. The leading-order exponent in its Painlev\'e series expansion arises as $p=0$. This indicates that  the resulting expansion may correspond to (a special case of) the Taylor expansion solution or to the possibility that it  must be modified by adding  logarithmic terms. See the original article \cite{Pickering1997} for a  discussion of this theme. A possible answer to this question is beyond the scope of the current article.

\section*{Acknowledgements} We thank the reviewers for their useful comments and suggestions. Thanks are also due to Prof. M.~Dunajski for bringing Ref. \cite{DunajskiPrzanowski} to our attention.


\end{document}